\documentclass[twocolumn,tighten,trackchanges,onecolappendix]{aastex62}
\hypersetup{linkcolor=red,citecolor=blue,filecolor=cyan,urlcolor=magenta}

\usepackage{url}
\shorttitle{Detection of Coronal Magnetic Activity in Nearby AGNs}
\shortauthors{Inoue, Doi}

\begin{document}
\title{Detection of Coronal Magnetic Activity in Nearby Active Supermassive Black Holes} 

\correspondingauthor{Yoshiyuki Inoue, Akihiro Doi}
\email{yoshiyuki.inoue@riken.jp}
\email{akihiro.doi@vsop.isas.jaxa.jp}

\author[0000-0002-7272-1136]{Yoshiyuki Inoue}
\affil{Interdisciplinary Theoretical \& Mathematical Science Program (iTHEMS), RIKEN, Saitama 351-0198, Japan}

\author{Akihiro Doi}
\affiliation{Institute of Space and Astronautical Science JAXA, 3-1-1 Yoshinodai, Chuo-ku, Sagamihara, Kanagawa 252-5210, Japan}
\affiliation{Department of Space and Astronautical Science, The Graduate University for Advanced Studies (SOKENDAI),3-1-1 Yoshinodai, Chuou-ku, Sagamihara, Kanagawa 252-5210}

%%%%%%%%%%
%%    Abstract    %%
%%%%%%%%%%
\begin{abstract}
Central supermassive black holes of active galactic nuclei host hot plasma with a temperature of $10^9$~K, namely coronae. Like the Sun, black hole coronae are theoretically believed to be heated by their magnetic activity, which have never been observed yet. Here we report the detection of coronal radio synchrotron emission from two nearby Seyfert galaxies using  the Atacama Large Millimeter/submillimeter Array, the Karl G. Jansky Very Large Array, and Australia Telescope Compact Array. The coronal magnetic field of both systems is estimated to be $\sim10$~Gauss on scales of $\sim40$~Schwarzschild radii from the central black holes. This magnetic field strength is weaker than the prediction from the magnetically heated corona scenario. We also find that coronae of Seyferts are composed of both thermal and non-thermal electrons. This may imply a possible contribution of Seyferts to the cosmic MeV gamma-ray background radiation.
\end{abstract}


%%%%%%%%%%%%
%%    Introduction    %%
%%%%%%%%%%%%
\section{Introduction}
\label{sec:intro}
Magnetic fields in the vicinity of supermassive black holes (SMBHs) are believed to be one of the most fundamental parameters governing the launch of relativistic jets \citep{blandford77} and generation of hot accretion disk coronae \citep{haardt91}. Various attempts have been made to measure magnetic fields at the center of active galactic nuclei \citep[AGNs;][]{gnedin2014,lopez2015,Marti-Vidal2015}. To date, however, magnetic fields in the inner accretion disk region at the scale of tens of Schwarzschild radii ($r_s$) have not been measured yet. 

In the radio centimeter (cm) or longer wavelength regime, synchrotron radiation from AGN jets and/or galactic cosmic-ray particles dominate the emission, while extended galactic thermal dust emission dominates the millimeter (mm) radiation. It is theoretically predicted that there also exists radio synchrotron emission arising from accreting hot plasma, namely corona, near central SMBHs appearing in the mm band due to the strong synchrotron self-absorption (SSA) effect by the coronal component itself \citep{DiMatteo1997,laor2008,Veledina2011,inoue2014,Raginski:2016}. 

Previous observations had presented inconclusive signs of a new component in the radio spectra of several Seyfert galaxies.  The putative new component -- so-called mm excess -- had been explored only in the short cm regime \citep{antonucci1988,barvainis1996} or a monochromatic mm band \citep{behar2015,Behar2018}.  Recently, a multi-band observation at 7 mm and longer wavelengths presented a possible sign for the existence of mm excess \citep{doi2016}.  Synchrotron emission from thermal hot plasma and free-free emission from surrounding ionized materials are found to be difficult to explain the putative mm-excess in terms of luminosity \citep{Field1993,laor2008,doi2016}. However, it is still inconclusive evidence because a paucity of multi-band data at shorter wavelengths. Investigation using conventional mm telescopes has been also hampered by the contamination of extended galactic emission.  Thus, the existence as well as physical origins of those excesses are still veiled in mystery.  High angular resolution and multi-band observations are required for the detection and understanding of compact coronal synchrotron emission \citep{inoue2014}.

By utilizing the Atacama Large Millimeter/submillimeter Array (ALMA), we have obtained new measurements for nearby radio-quiet Seyferts IC~4329A and NGC~985 at 12 frequency bands from 90~GHz up to 250~GHz. The observed radio emission is hardly contaminated by dust because of the sub-arcsec resolution of ALMA. We further analyzed archival cm data from the Karl G. Jansky Very Large Array (VLA) and Australia Telescope Compact Array (ATCA). In this paper, we report our new observational results.

This paper is organized as follows. In Section 2, we introduce our observations and data analysis. In Section 3, we present our results. Discussion and conclusion are given in Sections 4 and 5, respectively.

\section{Observations}
\label{sec:obs}

\subsection{Target objects}
IC~4329A is a southern Seyfert~1.2 galaxy at $z=0.0161$ \citep{wil91} whose central black hole mass is $\sim1.2\times10^8~M_\odot$ \citep{nik04,mar09}. The host galaxy is an edge-on spiral galaxy in a pair with IC~4329 separated by $\sim3$~arcmin. IC~4329A is one of the brightest X-ray Seyferts and shows a typical spectrum of radio-quiet Seyferts in the hard X-ray band \citep{zdz94,mad95,zdz96,brenneman2014a}. 

NGC~985 is a southern Seyfert~1 galaxy at $z=0.0427$ \citep{Fisher1995} whose black hole mass is $\sim2.2\times10^8~M_\odot$ \citep{kim2008}. NGC~985 is a peculiar galaxy with a prominent ring-shaped zone at several kiloparsecs from the nucleus \citep{deVaucouleurs1975}, suggesting that the galaxy is under- going a merging process. The galaxy contains a double nucleus, Seyfert nucleus, and non-AGN \citep{PerezGarcia1996}. The nuclei are separated by 3~arcsec.

\subsection{Radio observations with ALMA}
We observed IC~4329A and NGC~985 on October 2nd, 2016 and August 18th, 2017, respectively, using the ALMA Band~3, ~4, and 6 receivers (project-ID:  $\#$2016.1.01140.S, PI = Y. Inoue).  

For the IC~4329A observation, J1427-4206 and J1351-2912 were observed as a flux calibrator and phase calibrator, respectively.  Additionally, J1349-3056 was also observed for Band~6 as a water vapor radiometer calibrator.  
For the NGC~985 observation, J0238+1636 and J0241-0815 were observed as a flux calibrator and phase calibrator, respectively, for Bands 3 and 4, while J0334-4008 and J0241-0815 were observed as a flux calibrator and phase calibrator, respectively, for Band 6. Additionally, J0243-0550 was also observed as a water vapor radiometer calibrator.

Four spectral windows (SPWs) with a bandwidth of 2.000~GHz were placed at sky frequencies centered at 90.5~GHz, 92.4~GHz, 102.5~GHz, and 104.5~GHz in Band~3, at 138.0~GHz, 139.9~GHz, 150.0~GHz, and 152.0~GHz in Band~4, and at 213.0~GHz, 215.0~GHz, 229.0~GHz, and 231.0~GHz in Band~6. The correlation was made in the Time Division Mode with 128~channels for an SPW. 

The data were calibrated using the ALMA reduction package {\tt CASA} version 4.7.0 \citep{McMullin:2007}.  We reduced raw data using pipeline scripts provided by the ALMA regional center. 

For IC~4329A observation, the flux scaling factors were determined with reference to the quasar J1427-4206 at all bands.  For NGC~985 observations, the flux scaling factors were determined with reference to the quasar J0238+1636 at Bands 3 and 4, and J0334-4008 at Band~6.   
Their flux densities are regularly monitored by the ALMA observatory. We queried the ALMA calibrator database to make the assurance doubly sure.  J1427-4206 was stable ($\sim2.5~{\rm Jy}$ and $\sim1.2~{\rm Jy}$ at Bands~3 and 6, respectively) during the period before and after the dates of ALMA observations. The flux scaling factors used in the pipelines were based on a power-law spectrum that was determined from Band-3/7 observations carried out by the observatory within 7~days of our ALMA observations.  The flux densities of J0238+1636 and J0334-4008 were determined from Band-3/7 observations carried out by the observatory 2~days before our ALMA observations.   

We confirmed that no line emission practically contributes to the observed frequency ranges. Although a possible absorption feature by the Earth's atmosphere is seen at $\sim231.5~{\rm GHz}$ (spw~3 in Band~6), this does not affect the photometry of continuum emission. 

The calibrated visibilities were exported for four individual SPWs, and processed for mapping and CLEANing using natural weighting. Consequently, we obtained four continuum images for each band. The images were analyzed with the {\tt CASA}.  The ratio between peak flux density and integrated flux density was $\sim93$\%, $\sim92$\%, and $\sim85$\% for Band~3, 4, and 6, respectively, in the case of IC~4329A, while it was $\sim96$\%, $\sim90$\%, and $\sim90$\% in the case of NGC~985. Thus, unresolved sources dominate the emission in all the images. 

The emission center of the IC~4329A nucleus is located at $\alpha=$ 13h49m19.26170s $\pm 0.00262^{\prime\prime}$,   $\delta = -30^\circ$18$^\prime$34.22145$^{\prime\prime}\pm0.00264^{\prime\prime}$ (J2000) at 150~GHz.   
The emission center of the NGC~985 nucleus is located at  $\alpha=$ 02h34m37.84028s $\pm 0.00244^{\prime\prime}$, $\delta = -08^\circ$47$^\prime$16.07808$^{\prime\prime}\pm0.00173^{\prime\prime}$ (J2000) at 150~GHz.    
These are the best determined positions ever for these Seyfert nuclei where a systematic error is included. An astrometric position error on the phase calibrator was determined to be in a sub-milli-arcsecond level in the 2nd International Celestial Reference Frame \citep{Ma2009}. 

The flux measurements were performed by a Gaussian fit using the {\tt CASA} task {\tt IMFIT}. Results are summarized in Tables.~\ref{table:observation_ic4329a} and \ref{table:observation_ngc985}. We assume that the accuracy of the absolute amplitude calibration is 5\% for Bands 3 and 4, and 10\% for Band 6 following the ALMA Proposer's Guide. No other emission was found in a field of view of  $\sim 10~{\rm arcsec}$, even on higher signal-to-noise images that we made from visibilities concatenated throughout the four SPWs.

\begin{table*}
\caption{Spectral analysis results of ALMA, VLA, and ATCA observations of IC~4329A.\label{table:observation_ic4329a}} 
\begin{center}
\begin{footnotesize}
\begin{tabular}{ccccccccc}
\hline
\hline
$\nu$   &   $F_\nu$         &   $\sigma$   &   $\theta_\mathrm{maj} \times \theta_\mathrm{min}$         &   $PA$  & Date \\
(GHz)   &   (mJy)         &   (mJy/beam)   &   (arcsec$\times$arcsec)         &   (deg)   & (yr-m-d) \\
(1)   &   (2)   &   (3)         &   (4)   &   (5) & (6)\\
\hline
\multicolumn{6}{c}{ALMA}\\
\hline
% Band 3
90.5 & $8.00\pm0.49$ &  $0.094$ & $0.45  \times 0.44$  & $-25.3$ & 2016-10-02\\
92.4 & $7.92\pm0.49$  & $0.096$ & $0.44  \times 0.43$  & 24.6 & 2016-10-02 \\
102.5 & $8.07 \pm0.50 $  & $0.11$  & $0.43  \times 0.40$  & 20.0 & 2016-10-02 \\
104.5 & $7.81 \pm0.47 $  & $0.10$  & $0.42  \times 0.39$  & 23.7 & 2016-10-02 \\
% Band 4
138.0 & $6.69 \pm0.38 $  & $0.097$  & $0.28  \times 0.25$  & $-173.2$ & 2016-10-02 \\
139.9 & $6.79 \pm0.39 $  & $0.11$  & $0.27  \times 0.25$  & 18.9 & 2016-10-02 \\
150.0 & $6.45 \pm0.39 $  & $0.10$  & $0.29  \times 0.24$  & 26.0 & 2016-10-02 \\
152.0 & $6.27 \pm0.38 $  & $0.13$  & $0.25  \times 0.23$  & 9.6 & 2016-10-02\\
% Band 6
213.0 & $4.39 \pm1.21 $  & $0.16$  & $0.17  \times 0.15$  & 21.8 & 2016-10-02 \\
215.0 & $4.85 \pm0.84 $  & $0.15$ & $0.18  \times 0.16$  & 62.6 & 2016-10-02 \\
229.0 & $5.17 \pm0.92 $  & $0.15$  & $0.18  \times 0.14$  & 34.9 & 2016-10-02 \\
231.0 & $4.98 \pm0.94 $  & $0.16$ & $0.16  \times 0.14$  & 19.5 & 2016-10-02  \\
\hline
\multicolumn{6}{c}{VLA}\\
\hline
1.4	&$	66.8	\pm	4.7	$&$	2.10	$&$	45.0	\times	45.0	$&$	0.0	$&	1995-01-18	\\
1.4	&$	60.5	\pm	2.4	$&$	0.77	$&$	8.1	\times	3.6	$&$	48.1	$&	1998-02-08	\\
1.5	&$	63.5	\pm	4.8	$&$	2.20	$&$	45.0	\times	45.0	$&$	0.0	$&	1991-03-08	\\
4.9	&$	33.1	\pm	1.4	$&$	0.49	$&$	35.2	\times	10.1	$&$	11.6	$&	1991-03-08	\\
4.9	&$	19.5	\pm	0.8	$&$	0.25	$&$	1.3	\times	0.5	$&$	17.6	$&	1986-04-03	\\
8.4	&$	13.9	\pm	1.2	$&$	0.59	$&$	0.8	\times	0.4	$&$	-26.2	$&	1995-07-15	\\
8.4	&$	15.5	\pm	0.7	$&$	0.23	$&$	4.1	\times	1.8	$&$	-48.0	$&	1998-02-08	\\
\hline
\multicolumn{6}{c}{ATCA}\\
\hline
4.8	&$	28.5 	\pm	1.6 	$&$	0.65 	$&$	7.2	\times	2.5	$&$	9.6	$&	2008-04-19	\\
5.5	&$	34.6 	\pm	1.2 	$&$	0.31 	$&$	48.1	\times	30.6	$&$	69.3	$&	2010-03-04	\\
8.6	&$	17.8 	\pm	1.6 	$&$	0.77 	$&$	3.9	\times	1.4	$&$	9.6	$&	2008-04-19	\\
9.0	&$	24.4 	\pm	1.2 	$&$	0.45 	$&$	29.0	\times	18.5	$&$	69.3	$&	2010-03-04	\\
33.0	&$	14.3 	\pm	1.6 	$&$	0.32 	$&$	8.5	\times	5.0	$&$	-86.1	$&	2009-04-25	\\
35.0	&$	14.7 	\pm	1.7 	$&$	0.41 	$&$	8.0	\times	4.7	$&$	-86.1	$&	2009-04-25	\\
\hline
\end{tabular}
\end{footnotesize}
\end{center}
\begin{flushleft}
\begin{footnotesize}
Col.~(1) center frequency; 
Col.~(2) total flux density;   
Col.~(3) image r.m.s.~noise in the blank sky; 
Col.~(4)--(5) synthesized beam sizes along the major axis, minor axis, and position angle of the major axis, respectively;
Col.~(6) observation date.  
\end{footnotesize}
\end{flushleft}
\end{table*}

\begin{table*}
\caption{Spectral analysis results of ALMA, VLA, and ATCA observations of NGC~985.\label{table:observation_ngc985}}
\begin{center}
\begin{footnotesize}
\begin{tabular}{ccccccccc}
\hline
\hline
$\nu$   &   $F_\nu$         &   $\sigma$   &   $\theta_\mathrm{maj} \times \theta_\mathrm{min}$         &   $PA$  & Date \\
(GHz)   &   (mJy)         &   (mJy/beam)   &   (arcsec$\times$arcsec)         &   (deg)   & (yr-m-d) \\
(1)   &   (2)   &   (3)         &   (4)   &   (5) & (6)\\
\hline
\multicolumn{6}{c}{ALMA}\\
\hline
90.5	& $	1.72 	\pm	0.12 	$ & 	$0.058$ 	& $	0.25	\times	0.22	$ & $	87.7	$ &	2017-08-18	\\
92.4	& $	1.92 	\pm	0.13 	$ & 	$0.050$ 	& $	0.25	\times	0.21	$ & $	98.9	$ &	2017-08-18	\\
102.5	& $	1.64 	\pm	0.13 	$ & 	$0.059$ 	& $	0.23	\times	0.19	$ & $	84.7	$ &	2017-08-18	\\
104.5	& $	1.59 	\pm	0.13 	$ & 	$0.067$ 	& $	0.22	\times	0.19	$ & $	101.2	$ &	2017-08-18	\\
138.0	& $	1.71 	\pm	0.14 	$ & 	$0.058$ 	& $	0.17	\times	0.14	$ & $	100.0	$ &	2017-08-18	\\
139.9	& $	1.54 	\pm	0.12 	$ & 	$0.048$ 	& $	0.18	\times	0.15	$ & $	83.2	$ &	2017-08-18	\\
150.0	& $	1.44 	\pm	0.12 	$ & 	$0.052$ 	& $	0.16	\times	0.13	$ & $	99.0	$ &	2017-08-18	\\
151.8	& $	1.57 	\pm	0.12 	$ & 	$0.058$ 	& $	0.16	\times	0.13	$ & $	100.3	$ &	2017-08-18	\\
213.0	& $	1.35 	\pm	0.10 	$ & 	$0.041$ 	& $	0.12	\times	0.10	$ & $	84.4	$ &	2017-08-18	\\
215.0	& $	1.27 	\pm	0.10 	$ & 	$0.044$ 	& $	0.12	\times	0.09	$ & $	83.2	$ &	2017-08-18	\\
229.2	& $	1.26 	\pm	0.10 	$ & 	$0.040$ 	& $	0.11	\times	0.09	$ & $	85.6	$ &	2017-08-18	\\
231.0	& $	1.22 	\pm	0.09 	$ & 	$0.037$ 	& $	0.11	\times	0.09	$ & $	86.4	$ &	2017-08-18	\\
\hline
\multicolumn{6}{c}{VLA}\\
\hline
1.4	& $	13.1	\pm	1.0	$ & 	$0.45$	& $	45.0	\times	45.0	$ & $	0.0	$ &	1982-05-30	\\
4.9	& $	1.2	\pm	0.3	$ & 	$0.10$	& $	0.5	\times	0.4	$ & $	-18.7	$ &	1982-05-30	\\
14.9	& $	1.8	\pm	0.8	$ & 	$0.40$	& $	2.2	\times	1.5	$ & $	-14.9	$ &	1985-07-28	\\
1.5	& $	13.0	\pm	1.0	$ & 	$0.58$	& $	26.3	\times	15.3	$ & $	-33.1	$ &	1986-11-08	\\
4.9	& $	3.7	\pm	0.2	$ & 	$0.10$	& $	8.0	\times	4.6	$ & $	-31.1	$ &	1986-11-08	\\
1.5	& $	4.0	\pm	1.6	$ & 	$0.80$	& $	4.8	\times	2.8	$ & $	19.7	$ &	1990-05-23	\\
4.9	& $	1.6	\pm	0.7	$ & 	$0.24$	& $	1.4	\times	0.9	$ & $	10.3	$ &	1990-05-23	\\
8.4	& $	1.1	\pm	0.2	$ & 	$0.12$	& $	0.9	\times	0.7	$ & $	78.5	$ &	1990-05-23	\\
22.5	& $	1.9	\pm	0.3	$ & 	$1.68$	& $	3.8	\times	1.7	$ & $	72.3	$ &	2001-09-28	\\
8.5	& $	0.8	\pm	0.1	$ & 	$0.04$	& $	0.4	\times	0.2	$ & $	12.0	$ &	2003-06-19	\\
8.5	& $	1.3	\pm	0.1	$ & 	$0.10$	& $	1.7	\times	0.8	$ & $	-41.2	$ &	2003-12-24	\\
14.9	& $	0.8	\pm	0.3	$ & 	$0.25$	& $	0.9	\times	0.4	$ & $	-39.3	$ &	2003-12-24	\\
22.5	& $	1.2	\pm	0.3	$ & 	$0.13$	& $	0.5	\times	0.3	$ & $	-35.2	$ &	2003-12-24	\\
43.3	& $	3.1	\pm	1.2	$ & 	$0.62$	& $	0.3	\times	0.1	$ & $	-30.2	$ &	2003-12-24	\\
\hline
\multicolumn{6}{c}{ATCA}\\
\hline
19.0	& $	1.7 	\pm	0.7 	$ & 	$0.39$ 	& $	16.1	\times	8.6	$ & $	57.2	$ &	2009-10-15	\\
19.0	& $	1.7 	\pm	0.6 	$ & 	$0.38$ 	& $	17.7	\times	9.2	$ & $	-72.7	$ &	2009-10-16	\\
19.0	& $	2.0 	\pm	0.7 	$ & 	$0.39$ 	& $	15.8	\times	8.7	$ & $	57.3	$ &	2009-10-17	\\
21.0	& $	4.9 	\pm	1.9 	$ & 	$0.55$ 	& $	14.6	\times	7.8	$ & $	57.2	$ &	2009-10-15	\\
21.0	& $	2.3 	\pm	0.9 	$ & 	$0.22$ 	& $	16.0	\times	8.3	$ & $	-72.7	$ &	2009-10-16	\\
21.0	& $	1.5 	\pm	0.7 	$ & 	$0.39$ 	& $	14.3	\times	7.8	$ & $	57.3	$ &	2009-10-17	\\
\hline
\end{tabular}
\end{footnotesize}
\end{center}
\begin{flushleft}
\begin{footnotesize}
Same as in Table. \ref{table:observation_ic4329a}.
\end{footnotesize}
\end{flushleft}
\end{table*}

\subsection{Low-frequency radio data from VLA and ATCA observations}
We retrieved archival data obtained using the VLA and the ATCA from the US National Radio Astronomy Observatory and the Australia Telescope National Facility, respectively, to examine low-frequency components associated with the Seyfert nuclei.  For IC~4329A we retrieved all available data from the data archive centers; the observation codes were AE76, AM384, AS230, and AK394 for the VLA data and C1803 and C1964 for the ATCA data.  For NGC~985, we retrieved archival data with codes AB417 and AB489 for the VLA data and C1964 for the the ATCA data.  Additionally, we used the results of previous VLA studies for NGC~985 \citep{doi2016}.  We also retrieved and analyzed the 1.4-GHz images from the NRAO VLA Sky Survey~(NVSS; \citep{Condon:1998}). 

The VLA data were reduced in the standard manners for continuum observations using the software Astronomical Image Processing System~\citep[{\tt AIPS};][]{Greisen:2003}. Calibrations were carried out in accordance with the guidelines for the accurate flux density bootstrapping. Flux density bootstrapping covers the corrections for the elevation dependences of the gain curve for antenna elements and of atmospheric opacity. It also covers self-calibration procedures with the source structure models for standard flux calibrators and with a point source model for gain calibrators. 

The accuracy of flux scaling is 1\%--2\% at 8.4~GHz or lower in VLA observations; we assume a 3\% error in the present study, conservatively.  All target images were generated from calibrated visibilities with natural weighting in addition to deconvolution through the CLEAN algorithm. A single emission component was clearly detected at the position of the nucleus.  The source identification was performed by Gaussian model fitting on each image using the {\tt AIPS} task {\tt JMFIT}.  The final error of the flux density measurements was determined by root-sum-squares of the Gaussian fitting error and the flux scaling uncertainty. The fitting results were indicative of an underlying emission that was slightly resolved to be a few~arcsec in diameter on images with a compact array configuration.  In NGC 985, it is known that a radio emitting region of star formation was located $\sim17\arcsec$ westward from the nucleus \citep{Appleton2002}.  The NVSS image actually shows a westward elongation.  We made image-based model fitting with two components and subtracted the contribution of the western component from the total flux densities for the NVSS and the AB417 at 1.5~GHz.    

The ATCA data were also reduced in the standard manners using the software {\tt Miriad} \citep{Sault:1995}.  The ATCA observation and data analysis procedure also follows a philosophy that is very similar to the VLA observation. Corrections to atmospheric opacity, antenna positions, and system temperature measurements were applied to the data obtained at 19~GHz or higher frequencies.  We assume flux scaling errors of 10\% and 3\% at 33/35~GHz and other lower frequencies, respectively, according to the ATCA user guide.  Target images were synthesized with natural weighting and the CLEAN deconvolution.  Flux density measurements were carried out in the same manner as in the VLA data analysis, using the {\tt AIPS} task {\tt JMFIT}.  The VLA and ATCA results are listed in Tables~\ref{table:observation_ic4329a} and \ref{table:observation_ngc985}.

% stacking on NGC~985 ATCA data  --> figure caption

\subsection{X-ray data analysis of NGC~985}
\label{app:xray}
For the thermal coronal properties of IC~4329A, we adopted the data analysis results of the joint {\it Suzaku} and {\it NuSTAR} observations toward IC~4329A  \citep{brenneman2014b}. They found that the coronal temperature and the Thomson scattering optical depth are 50~keV and 2.34, respectively, assuming a spherical geometry. For NGC~985, we analyze archival {\it NuSTAR} data in order to derive coronal properties. Coronal parameters are determined by characteristics of primary high-energy continuum. Therefore, we restrict our analysis to 5--79~keV in order to avoid the effects of low-energy absorption, which are poorly constrained by {\it NuSTAR} only.

\begin{figure}
 \begin{center}
  \includegraphics[width=7.5cm]{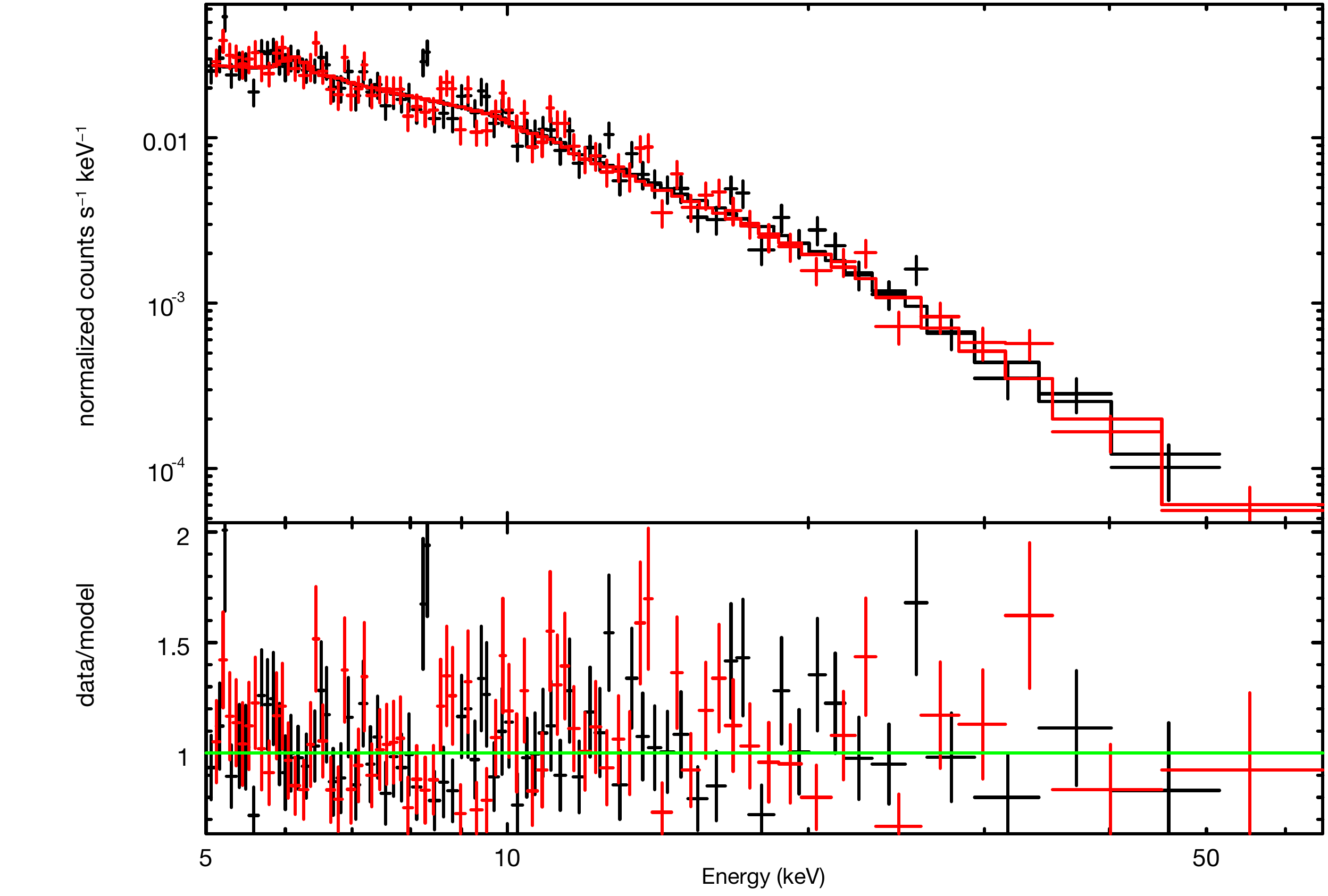} 
 \end{center}
\caption{Upper panel: The {\tt compTT} + {\tt pexmon}  spectral model fit in the range of 5--79~keV to the FPMA (black) and FPMB (red) data. Lower panel: data-to-model ratio for the FPMA (black) and FPMB (red). The horizontal line shows a ratio of unity for reference.
}\label{fig:xray_ngc985}
\end{figure}

{\it NuSTAR} is a focusing hard X-ray telescope providing a bandpass of 3--79~keV with a spectral resolution of $\sim1$~keV \citep{har13}. {\it NuSTAR} observed NGC~985 on August 11, 2013 (obsid: 60061025002). After eliminating Earth occultations, passages through the South Atlantic Anomaly, and other periods of high background, the NuSTAR observation reached $\sim13.9$ ks of on-source time for focal plane module A (FPMA) and focal plane module B (FPMB). 

We reduced the data using the NuSTAR Data Analysis Software (NuSTARDAS) and calibration version 1.4.1. We filtered the event files and applied the default depth correction using the nupipeline task. We used circular extraction regions, 45 arcsec in radius, for the source and background, with the source region centered on NGC~985 and the background taken from the corner of the same detector, as close as possible to the source without being contaminated by the point spread function wings. Spectra and images were extracted and response files were generated using the nuproducts task. To minimize systematic effects, we fitted spectra from FPMA and FPMB simultaneously. We allowed the absolute normalization for both modules to vary. We found a cross-calibration factor of $1.05^{+0.45}_{-0.23}$ for FPMB relative to FPMA.

Given the short exposure, we followed the analysis procedure of IC~4329A \citep{brenneman2014a}. we first evaluated the {\it NuSTAR} spectrum with the {\tt pexmon} model \citep{Nandra2007}, which incorporates both primary emission and reflection. We fixed the abundance of elements to the solar abundances, the inclination angle of $i=60^\circ$, otherwise we cannot get any constraints on other parameters. The foreground Galactic column density was set to $N_{\rm H} = 3.17\times 10^{20}$~${\rm cm^{-2}}$ \citep{Kalberla2005}. We got $\Gamma=1.71\pm0.08$, $E_{\rm cut}=99_{-27}^{+19}~{\rm keV}$, and $R=0.80_{-0.14}^{+0.31}$ with a goodness-of-fit of $\chi^2/\nu=191/145 (1.32)$.

Then, to derive physical parameter of the corona, we employed the {\tt compTT} model \citep{Titarchuk1994}, which calculates the Comptonization spectra. {\tt compTT} does not self-consistently calculate the reflection component. Thus, we added the reflection component by using {\tt pexmon}.  We set the cutoff energy in {\tt pexmon} to 2.7 times of the coronal electron temperature. We fixed the other values of the reflection component to the values found for the {\tt pexmon} fit and include only the reflection part of the model. We found the coronal temperature of $kT_e=29_{-9}^{+5}~{\rm keV}$ and the Thomson scattering optical depth of $\tau=3.5_{-0.6}^{+1.2}$ with a goodness-of-fit of $\chi^2/\nu=191/146 (1.31)$. In order to obtain the better fit, we may need further exposure. The resulting spectrum is shown in Figure. \ref{fig:xray_ngc985}.

\begin{figure}
 \begin{center}
  \includegraphics[width=7.5cm]{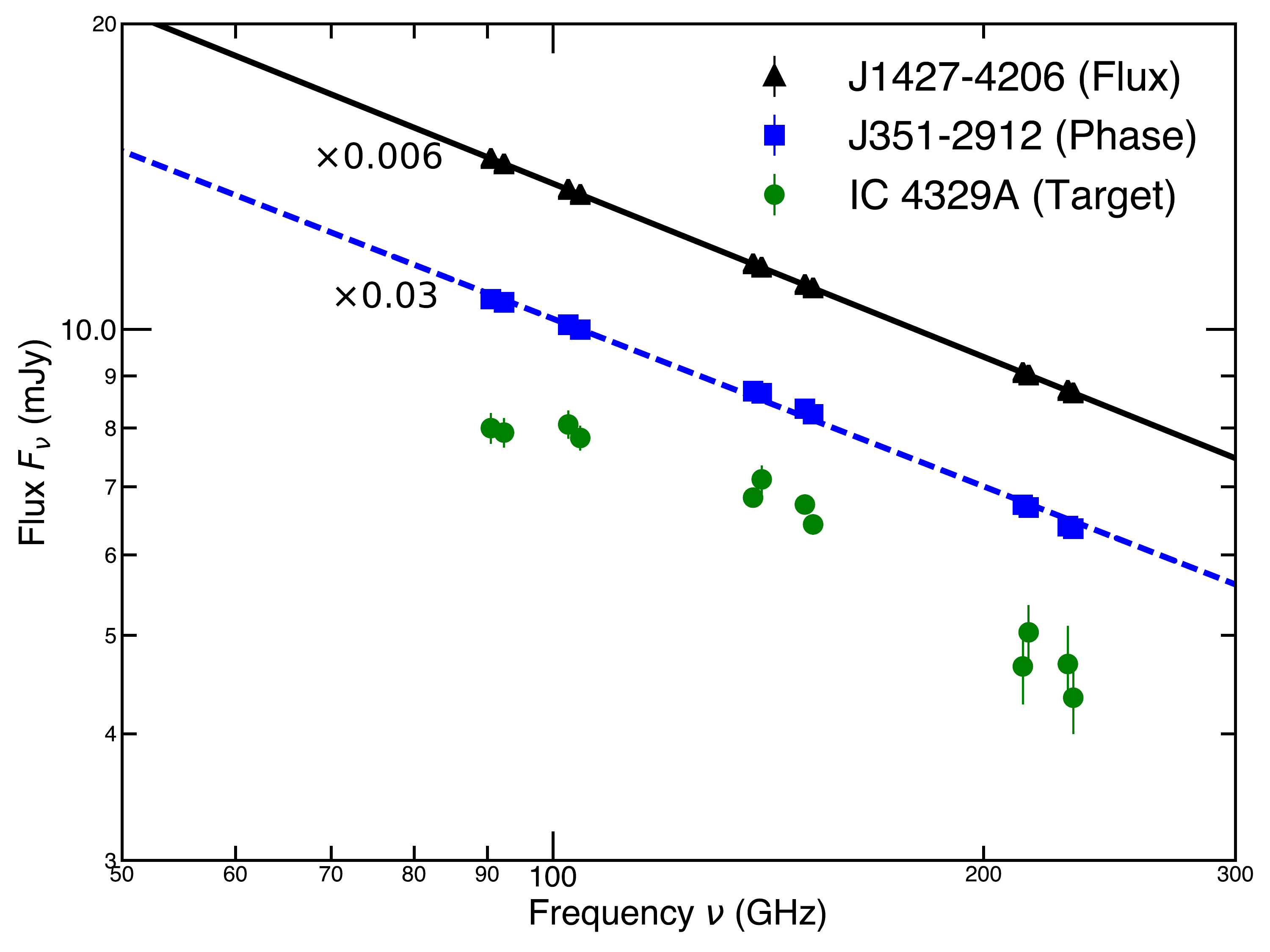} 
 \end{center}
\caption{The mm spectra of J1427-4206 (Flux calibrator, triangle), J1351-2912 (Phase calibrator, square), and IC~4329A (Target, circle) taken by ALMA. Each data point shows its 1-$\sigma$ error bar. Data points of J1427-4206 and J1351-2912 are renormalized by a factor of 0.006 and 0.03, respectively. The solid and dashed lines correspond to the single power-law fit to J1428-4206 and J1351-2912, respectively.}\label{fig:calibrator}
\end{figure}

\begin{figure*}
 \begin{center}
  \includegraphics[width=8.9cm]{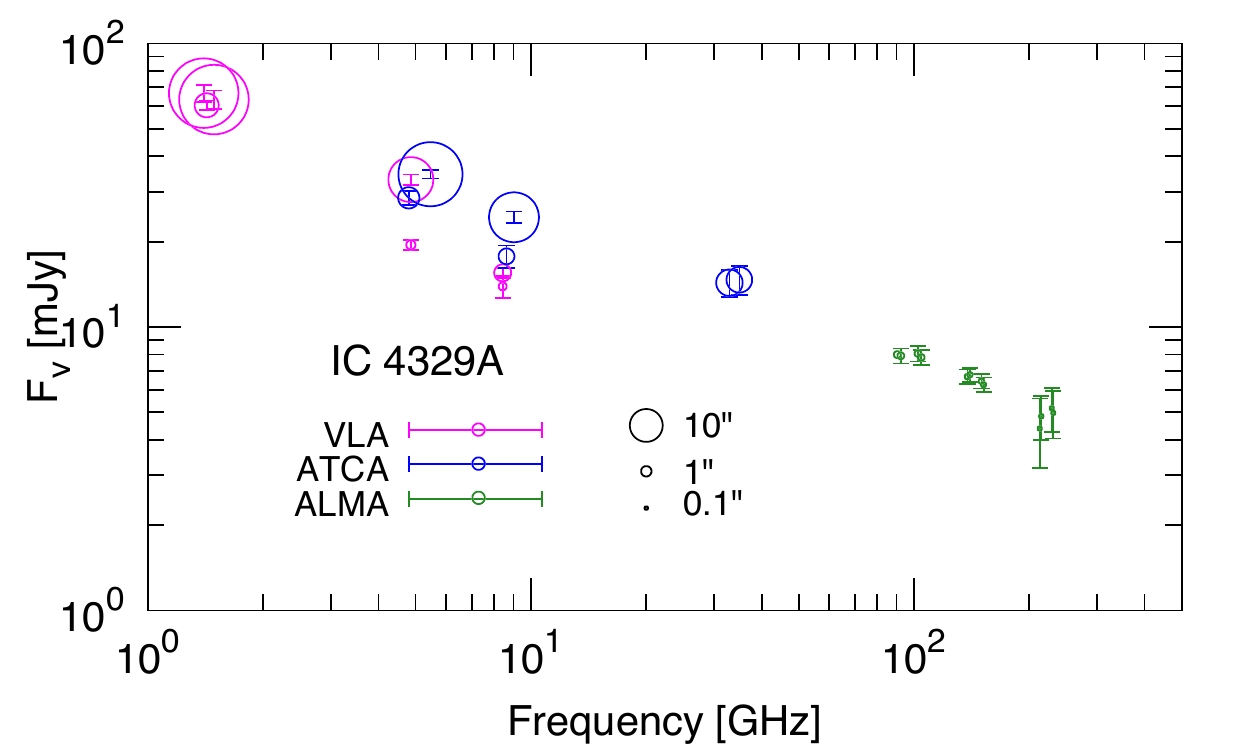} 
  \includegraphics[width=8.9cm]{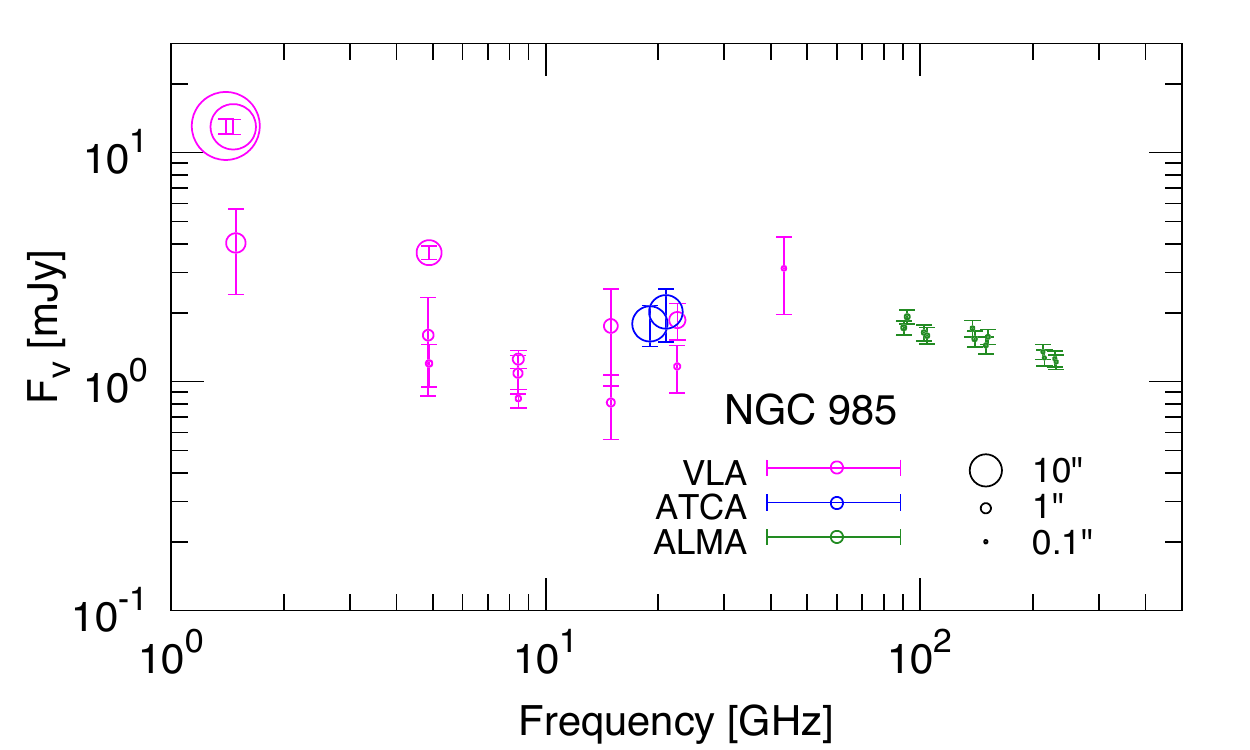} 
 \end{center}
\caption{{\it Left}: The cm-mm spectrum data of IC~4329A. The magenta, blue, and green points show the VLA, ATCA, and ALMA data, respectively. The error bars correspond to 1-$\sigma$ uncertainties. A circle around each data point represents the scaled synthesized beam size. The beam size dependency is presented as circles in the panel for the cases of $10\arcsec$, $1\arcsec$, and $0\farcs1$.  {\it Right}: Same as the left panel but for NGC~985.  The ATCA measurements at 19 and 21~GHz were averaged throughout three consecutive days.}\label{fig:SED_radio_raw}
\end{figure*}

\section{Results}
\label{sec:result}
Figure.~\ref{fig:calibrator} shows the mm spectrum of our target object IC~4329A together with those of the flux calibrator J1427-4206 and the phase calibrator J1351-2912. The flux of J1427-4206 and J1351-2912 in Figure.~\ref{fig:calibrator} is renormalized by a factor of 0.006 and 0.03, respectively, for the comparison with the target object IC~4329A. The spectra of J1427-4206 and J1351-2912 are well reproduced by single power-law models, while our target object IC~4329A shows spectral curvature (Figure. \ref{fig:calibrator}). We do not see such curvature features in the ALMA data of NGC~985.

The left and right panel of Figure. \ref{fig:SED_radio_raw} shows the observed cm-mm radio spectral data of IC~4329A and NGC~985, respectively. Different beam size data are included ranging from 0.15" to  45.0" for IC~4329A and from 0.10" to  45.0" for NGC~985 (Tables~\ref{table:observation_ic4329a} and \ref{table:observation_ngc985}). Geometric mean of synthesized beam sizes at major and minor axes for each data is also represented in the Figure. In both objects, a clear tendency of higher flux densities at larger beam sizes is apparent. For large beam size data sets, it is naturally expected that extended emission such as host galaxy component contributes to photometry, while such extended emission does not significantly contribute to small beam size data sets. Therefore, consideration of observational beam size is important for the understanding the emission from the compact regions.

\begin{figure*}
 \begin{center}
  \includegraphics[width=8.9cm]{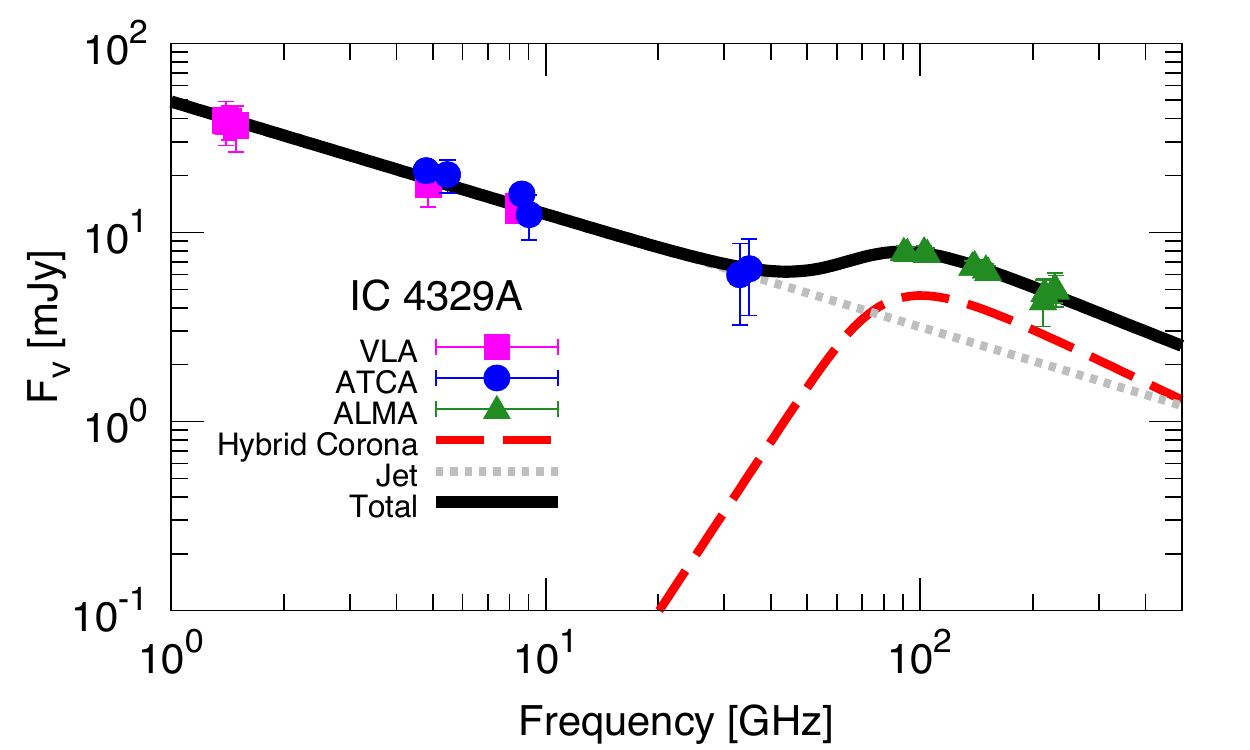} 
  \includegraphics[width=8.9cm]{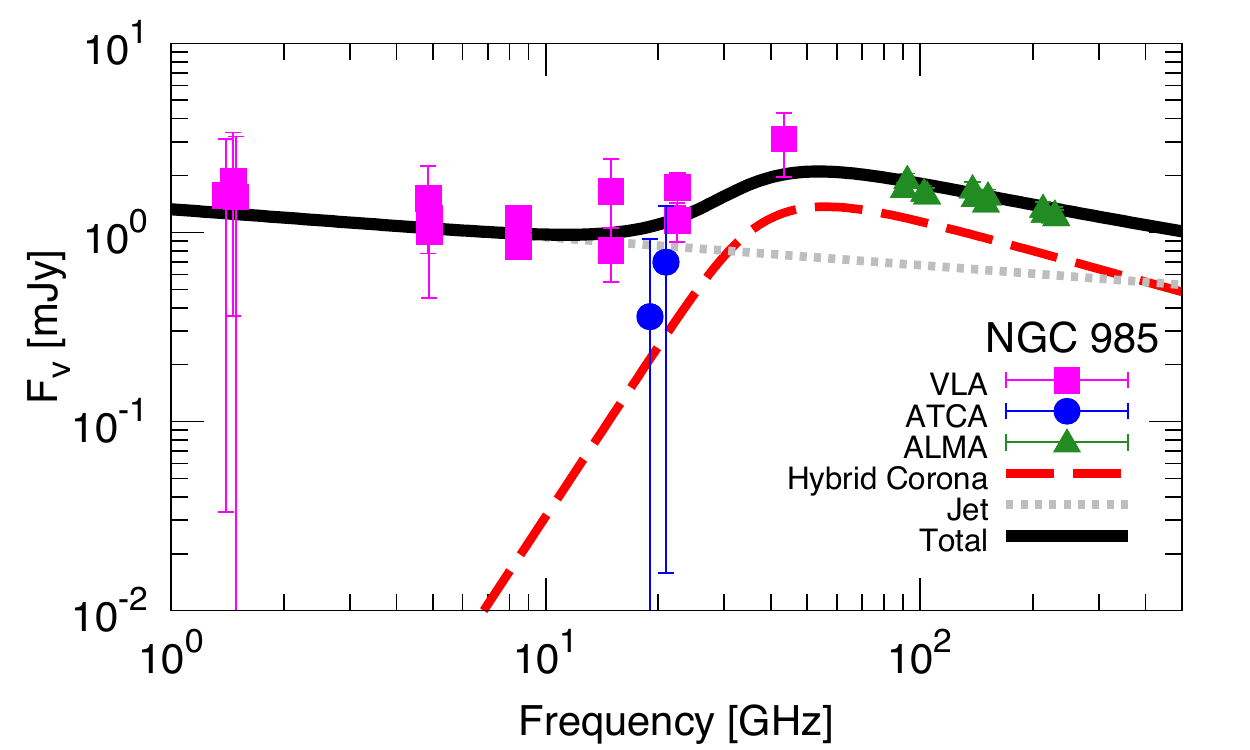} 
 \end{center}
\caption{{\it Left}: The cm-mm spectrum of IC~4329A after subtracting extended emission due to galactic star formation activity. The square, circle, and triangle points show the VLA, ATCA, and ALMA data, respectively. The error bars correspond to 1-$\sigma$ uncertainties where we take into account flux measurement uncertainties and galactic component model uncertainties. The dashed and dotted lines show the fitted hybrid corona and jet component, respectively, with parameters shown in Table.~\ref{table:spectral_data}. The solid line shows the sum of these two components. {\it Right}: Same as the left panel but for NGC~985.  The ATCA measurements at 19 and 21~GHz were averaged throughout three consecutive days.}\label{fig:SED_radio}
\end{figure*}

In order to investigate the origin of the radio emission, we simultaneously fit ALMA, VLA, and ATCA measurements between 1~GHz and 250~GHz with four spectral components. One is synchrotron emission from a hybrid corona \citep{inoue2014} in which fitting parameters are the plasma size ($R$), the magnetic field strength ($B$), the spectral index of non-thermal electrons ($p$), and  the energy fraction of non-thermal electrons ($\eta$). The coronal temperature and the Thomson scattering optical depth are determined by X-ray spectral analysis  \citep[e.g.,][]{brenneman2014b}. We adopt spherical geometry for coronae for simplicity. We define the distribution of electrons in hybrid coronae as relativistic Maxwellian distribution at $\gamma_e \le \gamma_{\rm br}$ and $n_e(\gamma_e)\propto({\gamma_e}/{\gamma_{\rm br}})^{-p} $ at $ (\gamma_e > \gamma_{\rm br})$, where $\gamma_e$ is the electron Lorentz factor and $\gamma_{\rm br}$ is the break Lorentz factor. By setting $\eta$, $\gamma_{\rm br}$ can be obtained. Second is a jet synchrotron emission component represented by the form $F(\nu)=A_{\rm Jet}(\nu/\nu_0)^{\alpha_{\rm Jet}}$ setting $\nu_0=100~{\rm GHz}$. The other two components represent galactic thermal bremsstrahlung and galactic synchrotron of the form $F(\nu, \theta_{\rm maj}, \theta_{\rm min})=A_{\rm SF, FF}(\nu/\nu_0)^{-0.1} \allowbreak \min(1.0, \theta_{\rm maj}\theta_{\rm min}/\theta_{\rm SF}^2)$ and $A_{\rm SF, Syn}\allowbreak(\nu/\nu_0)^{-0.8}\allowbreak\min(1.0, \theta_{\rm maj}\theta_{\rm min}/\theta_{\rm SF}^2)$, where $\theta_{\rm maj}$ is the beam size along the major axis, $\theta_{\rm min}$ is that along the minor axis, and $\theta_{\rm SF}$ gives the angular size of the star formation activity region around the Seyfert nuclei. We set a spectral index of $0.1$ for the thermal free-free component and the spectral index of $0.8$ for the galactic synchrotron component \citep{Magnelli2015}. Since the angular resolutions of observations by VLA and ATCA were not as good as that of ALMA (Figure. \ref{fig:SED_radio_raw}), the measured fluxes by VLA and ATCA must have been contaminated not only by the jet but also by extended emission due to star formation activity in the host galaxy. Therefore, we consider beam sizes and $\theta_{\rm SF}$ in our modeling.

%The normalization can be determined from the Thomson scattering opacity $\tau$, which is determined by the thermal Comptonized X-ray spectrum.

The left panel of Figure. \ref{fig:SED_radio} shows the cm-mm radio spectrum of IC~4329A after we subtract beam-size dependent extended galactic emission components. The mm excess is clearly visible in the cm--mm spectrum, which determines the properties of the coronae. The measured ALMA spectrum of IC~4329A exhibits flattening below 100~GHz, while calibrators' spectra exhibit simple power-law shape (Figure. \ref{fig:calibrator}). This indicates SSA in a very compact region in the order of several tens of the Schwarzschild radius. The measured spectra are well reproduced by the assumed models without time variability. 

The coronal properties of IC~4329A are estimated as $R=39\pm10~r_s$ with $B=9\pm6$~G and $p=2.9\pm0.9$ based on our model fitting. The coronal temperature and the Thomson scattering optical depth are set to 50~keV and 2.34, respectively \citep{brenneman2014b}. We fix the energy fraction $\eta$ of non-thermal electrons as 4\% of that of the total electron population. This value is inferred from the explanation of the cosmic MeV gamma-ray background radiation by Seyferts \citep{Inoue2008}. Because $\eta$, $B$, and $R$ are closely tied \citep{inoue2014}, current radio data do not allow us to solve these three parameters simultaneously without decoupling thermal and non-thermal components. Future simultaneous radio and sub-MeV gamma-ray observations will enable us to estimate more precise model parameters. Even if we vary $\eta$ from 0.01 to 0.30, the parameters $B$, $R$, and $p$ are consistent within an uncertainty of 1-$\sigma$. Current X-ray observations of Seyferts constrain $\eta<0.3$ \citep{fabian2017}.

The right panel of Figure. \ref{fig:SED_radio} shows the cm-mm radio spectrum of NGC~985 after subtracting extended galactic components, same as in IC~4329A. Again, the mm excess is clearly seen. We fit the spectrum with the same models for IC~4329A. For the X-ray coronal properties, we adopt the coronal temperature and the optical depth due to Thomson scattering of 29~keV and 3.5, respectively (see Section. \ref{app:xray}). The coronal properties of NGC~985 are estimated as $(R, B, p) = (47\pm21~r_s, 4\pm2~{\rm G}, 2.1\pm0.3)$, setting $\eta=0.04$. The obtained parameters for both systems are summarized in Table. \ref{table:spectral_data}.

\begin{table}
\caption{Spectral fitting results in the frequency range of 1~GHz to 250~GHz.\label{table:spectral_data}}
\begin{center}
\begin{footnotesize}
\begin{tabular}{crr}
\hline
\hline
Parameter & IC~4329A & NGC~985 \\
\hline
\multicolumn{3}{c}{Hybrid corona}\\
\hline
$p$ & $2.90\pm0.86$ & $2.11\pm0.28$\\
$B$ (G) & $9.42\pm6.16$ & $4.18\pm 1.90$\\
$R$ ($r_s$) & $38.8\pm9.7$	 & $46.9\pm20.5$	\\
$\eta$ & $0.04$ (fixed) & $0.04$ (fixed)\\
\hline
\multicolumn{3}{c}{Jet}\\
\hline
$A_{\rm Jet}$ (mJy) & $3.23\pm0.79$ & $0.61\pm0.20$\\
$\alpha_{\rm Jet}$& $-0.59\pm0.09$& $-0.18\pm0.13$\\
\hline
\multicolumn{3}{c}{Star formation activity}\\
\hline
$A_{\rm SF, Syn}$ (mJy) & $0.60\pm0.28$ & $0.38\pm0.03$\\
$A_{\rm SF, FF}$ (mJy) & $6.16\pm1.92$ & $<0.35$\\
$\theta_{\rm SF}$ (arcsec)& $6.05\pm0.39$ & $7.69\pm0.47$\\
\hline
$\chi^2/{\rm d.o.f}$ & $11.3/17$ & $22.5/20$\\
\hline
\end{tabular}
\end{footnotesize}
\end{center}
\begin{flushleft}
\begin{footnotesize}
The errors and the upper limit correspond to 1-$\sigma$ uncertainties. The upper limit is the 1-$\sigma$ upper limit.
\end{footnotesize}
\end{flushleft}
\end{table}

\section{Discussion}
\label{sec:dis}

\subsection{Variability}
Both IC~4329A and NGC~985 are known to be relatively steady in X-ray ($\sim50$~\% of flux variability) during  {\it NuSTAR} and ALMA observations \citep{Oh2018,Kawamuro2018}. Such small variabilities will vary $B$ and $R$ for about 10\%, which is smaller than 1-$\sigma$ uncertainties of parameters. Here, compact jets explaining the cm data are also  potentially variable. But, the multi-epoch cm data are consistent with a single power law, indicating no significant variability in the jet components in both Seyferts.

\subsection{Contribution of Other Components}
Thermal dust emission might contribute to the mm flux \citep[e.g.][]{doi2016}. However, given the ALMA spectral shapes, the dust scenario is clearly disfavored because the gray body spectrum from thermal dust shows a much harder spectrum. IC~4329A is known to have a hot nuclear dust component \citep{Mehdipour2018}. The expected compact nuclear dust contamination into the ALMA bands is less than $1.6\times10^{-2}$~mJy.

Would a pure thermal coronal synchrotron emission scenario explain the observed spectra \citep{inoue2014,Raginski:2016}? Considering the X-ray measurements (temperature and Thomson scattering optical depth), the required size for pure thermal coronae become $R\sim2600 r_s$ with $B\sim330$~G for IC~4329A and $R\sim3100 r_s$ with $B\sim180$~G for NGC~985. Such large coronal size is naturally expected for pure thermal synchrotron cases \citep{inoue2014,Raginski:2016}. However, the inferred coronal size are too large comparing to the expected coronal size from other measurements. For example, a combined optical and X-ray study of 51 unobscured AGNs revealed the average corona radii of $(27\pm18)r_s$ \citep{Jin2012}. Microlensing observations toward the Lensed Quasar Q J0158-4325 also inferred the half-light radius for X-ray emission of $\sim10r_s$ \citep{morgan2012}. Given these size constraints, pure thermal electron distribution would be difficult to explain the measured radio spectral data of the two Seyferts.

Here, the SSA peak flux ($\sim10$~mJy) and frequency ($\sim100$~GHz) position are observationally fixed, while the number of high energy electron ($\gamma_e\gg1$) in pure thermal distribution is much less than that in hybrid distribution cases. Thus, the required size and magnetic field strength for thermal models become larger than those in the case of hybrid coronae. We note that the size $R$ and magnetic field $B$ of plasma can also be derived once the SSA frequency and flux are determined. The SSA optical depth is defined as $\tau_{\rm SSA}(\nu)=\alpha_{{\rm Syn}, \nu}R$ where $\alpha_{{\rm Syn}, \nu}$ is the synchrotron absorption coefficient. The SSA frequency $\nu_{\rm SSA}$ is given by $\tau_{\rm SSA}(\nu_{\rm SSA})=1$. The optically thin synchrotron flux is expressed as $F_{\rm Syn}=j_{\rm Syn, \nu} V / 4\pi d^2$ where $j_{\rm Syn, \nu}$ is the synchrotron emissivity, $V=4\pi R^3/3$ is the volume of the plasma, and $d$ is the distance to the object. By adopting synchrotron absorption and emission coefficients \citep{rybicki1979} and setting electron distribution, one can derive $R$ and $B$ by solving above equations.

\subsection{Corona Heating Mechanism and Its Structure}

It is believed that accretion disk coronae are heated by the release of magnetic field energy, like in the Sun, while cooling is dominated by Comptonization by disk photons in the case of disk coronae \citep{haardt91,liu2002}. Based on our measurement, the plasma beta $\beta$ (the ratio of gas pressure to magnetic pressure, $p_{\rm gas}/p_{\rm B}=8\pi nk_BT/B^2$) in the accretion disk coronae becomes $\sim0.1$ for both objects. The inferred plasma beta is higher than that typically observed in the solar corona, $10^{-3}$--$10^{-1}$ \citep{Gary2001}. Given our measurements of two Seyferts, the magnetic heating rate in the coronae is estimated as $Q_{\rm B, heat}\sim10^{10}-10^{11}~{\rm erg~cm^{-2}~s^{-1}}$ (i.e., magnetic energy release rate at the Alfv\'en speed), while the Compton cooling rate by disk photons is $Q_{\rm IC, cool}\sim10^{14}~{\rm erg~cm^{-2}~s^{-1}}$ setting 10\% of the Eddington luminosity. Therefore, magnetic field energy is insufficient for keeping coronae as hot as observed in X-rays.

Here, if the inner part of the accretion disk at several tens of the Schwarzschild radius is composed of a cold disk and a hot accretion flow like an advection-dominated accretion flow \citep[e.g.,][]{yamada2013}, the advection-heated inner accretion region can play the role of coronae \citep{Merloni2000,Kawabata2010}. Observations of a Galactic X-ray binary Cygnus X-1 suggests such an accretion disk picture in order to explain its X-ray spectra and its variability \citep{yamada2013,Basak2017}. In this case, the hot accretion flow would be heated up to electron rest-mass energy \citep{manmoto1997}. Based on the self-similar solutions of hot accretion flows \citep{kato2008,yuan2014}, the magnetic field strength can be estimated as $\sim10$~G at the size of $40r_s$ assuming the $\alpha$ parameter of 0.1 and the mass accretion rate of 0.1\% of the Eddington rate in the inner hot accretion flow. Following the numerical general relativistic magnetohydrodynamic simulations of hot accretion flows, $\beta$ decreases with increasing distance away from the midplane of the disk and becomes $\beta\sim0.1$ above approximately two density scale heights \citep{DeVilliers2003,DeVilliers2005}.  These are consistent with our radio observation results. 

Typically, seyferts have much higher accretion rate and the accretion rate  can be determined with reasonable accuracy from the observed optical flux once the black hole mass is known \citep[e.g.,][]{Davis2011}. However, optical emission arises at the outer thin disc region, i.e. the inferred high mass accretion rate is the value at the outer region. The accurate mass accretion rate in the inner region is still not well understood because AGN disk emission peaks in the UV regime which cannot be directly observed due to dust extinction. Moreover, \citet{brenneman2014b} found the equivalent width of the broad iron Fe~K$\alpha$ line emission is 24--42~eV corresponding to the location of cold material at $35 r_s$. This suggests that the optically thick disk may not extend down to the inner region. Therefore, the accretion rate in the inner region could be lower than that measured from optical fluxes. 

This cold + hot accretion flow scenario may also present an inconsistency with the general picture of broad iron Fe~K$\alpha$ line emissions originating from near the innermost stable circular orbit and soft X-ray lags  \citep{Reynolds2014,Kara2016}. However, line profiles of Fe~K$\alpha$ lines in Seyferts could be narrower than ever thought, implying the far location of neutral iron, by considering hard X-ray variabilities \citep{Noda2011} or double partial covering materials \citep{Miyakawa2012}. For the IC~4329A, as described above, the location of the origin of the Fe~K$\alpha$ line is found to be at $\sim35r_s$ \citep{brenneman2014b}. The origin of soft X-ray lags is also under debate \citep{Miller2010,Mizumoto2018}.  Future detailed X-ray and radio observations will be important to reconcile these problems.

\subsection{On the Origin of the Cosmic MeV Gamma-ray Background Radiation}
The possible evidence of non-thermal electrons in coronae would shed new light on the origin of the cosmic MeV gamma-ray background, which has been long discussed \citep[see e.g.,][for reviews]{inoue2014_Fermi}. One possible candidate is the MeV non-thermal photon spectral tail expected in Seyfert spectra arisen from non-thermal electrons in coronae \citep[see][for details]{Inoue2008}, while blazars are also considered as a candidate \citep{Ajello2009}. Due to a lack of MeV gamma-ray observations, the origin of the cosmic MeV gamma-ray background is still veiled in mystery, although anisotropy measurements may distinguish these two scenarios \citep{Inoue2013}. 

Our radio observations toward nearby Seyferts suggested possible existence of non-thermal electrons in coronae, which is anticipated for the Seyfert scenario on the MeV gamma-ray background. The required non-thermal electron energy fraction for the Seyfert scenario is approximately $4$\% with a spectral index of $\sim3.5$. Future simultaneous radio and MeV gamma-ray observations will be able to quantify amount of non-thermal electrons in coronae and reveal the origin of the MeV background radiation.

\section{Conclusion}
\label{sec:Conclusion}
By utilizing ALMA, we have obtained new measurements for nearby radio-quiet Seyferts IC~4329A and NGC~985 at 12 frequency bands from 90~GHz up to 250~GHz. We further analyzed archival cm data from VLA and ATCA. By combining the mm and cm data sets, the spectra of compact components clearly show the excesses in the mm band. 

The measured spectra suggest that the coronal magnetic field is approximately $10$~Gauss on scales of $\sim40r_s$ from the SMBHs for the two Seyferts. This magnetic field strength is weaker than the prediction from the magnetically heated accretion corona scenario \citep[e.g.,][]{liu2002}.  Our mm observation also provides observational evidence of non-thermal electron population in coronae, which may suggest possible contribution of Seyferts to the cosmic MeV gamma-ray background radiation through the Comptonization process of such non-thermal electrons \citep{Inoue2008}.

\acknowledgments
YI is supported by JSPS KAKENHI Grant Number JP16K13813, program of Leading Initiative for Excellent Young Researchers, MEXT, Japan, and RIKEN iTHEMS Program. This paper makes use of the following ALMA data: ADS/JAO. ALMA$\# 2016.1.01140.S$. ALMA is a partnership of ESO (representing its member states), NSF (USA), and NINS (Japan), together with NRC (Canada), MOST and ASIAA (Taiwan), and KASI (Republic of Korea), in cooperation with the Republic of Chile. The Joint ALMA Observatory is operated by ESO, AUI/NRAO, and NAOJ. This paper also utilizes archival data from VLA operated by NRAO. The National Radio Astronomy Observatory is a facility of the National Science Foundation operated under cooperative agreement by Associated Universities, Inc. This paper makes use of archived data obtained through the Australia Telescope Online Archive (\url{http://atoa.atnf.csiro.au}). The Australia Telescope Compact Array is part of the Australia Telescope National Facility, which is funded by the Australian Government for operation as a National Facility managed by CSIRO.

\vspace{5mm}
\facilities{ALMA, ATCA, NuSTAR, VLA}

\end{document}